\documentstyle[12pt,epsf,epsfig,wrapfig]{article}

\def\text{\rm }

\begin{document}
\vspace{-0.5cm}
\begin{flushright}
Jagellonian University Preprint TPJU-9/2001
\end{flushright}
\begin{center}
{\bfseries SU(3) SYMMETRY BREAKING AND POLARIZED PARTON DENSITIES}\footnote{
Presented at Dubna Workshop Spin 2001} 
\vskip 5mm
\underline{M. Prasza{\l}owicz}$^{1}$, H.-Ch. Kim$^{2}$ and K. Goeke$%
^{2} $ \vskip 5mm {\small (1) {\it M. Smoluchowski Institute of Physics,
Jagellonian University, \\[0pt]
ul. Reymonta 4, 30-059 Krak{\'o}w, Poland } \\[0pt]
(2) {\it Department of Physics, Pusan National University,\\[0pt]
Pusan 609-735, Republic of Korea } \\[0pt]
(2) {\it Institute for Theoretical Physics II, Ruhr-University Bochum, \\[0pt]
D-44780 Bochum, Germany } \\[0pt]
}
\end{center}


\begin{abstract}
We analyze the semileptonic weak decays of the octet baryons in a "{\em %
model independent}" approach, based on the algebraic structure of the Chiral
Quark-Soliton Model. We argue that this analysis is in fact more general
than the model itself. While the symmetry breaking for the semileptonic
decays themselves is not strong, other quantities like $\Delta s$ and $%
\Delta \Sigma$ are much more affected. We calculate $\Delta \Sigma$ and $%
\Delta q$ for all octet baryons. Unfortunately, large experimental errors of
$\Xi^-$ decays propagate in our analysis, in particular, in the case of $%
\Delta\Sigma $ and $\Delta s$. Only if the errors for these decays are
reduced, the accurate theoretical predictions for $\Delta\Sigma$ and $\Delta
s$ will be possible.
\end{abstract}


The experimental results on the first moment $I_{{\rm {p}}}$ of the proton
spin structure function $g_{1}^{p}$~\cite{EMC}--\nocite{SMC,E143,E154}\cite
{HERMES1}
\begin{equation}
I_{{\rm {p}}}=\int\limits_{0}^{1}dx\,g_{1}^{{\rm {p}}}(x)=\frac{1}{18}\left(
4\Delta u_{{\rm {p}}}+\Delta d_{{\rm {p}}}+\Delta s_{{\rm {p}}}\right)
\left( 1-\frac{\alpha _{{\rm {s}}}}{\pi }+\ldots \right) .  \label{Ip0}
\end{equation}
are usually interpreted in terms of the exact SU(3) symmetry. Then, in
contrast to the Ellis-Jaffe sum rule~\cite{EllisJaffe}, the strange quark
contribution to the nucleon spin deviates from zero and is of the order of $%
\Delta s=-0.1$, and the singlet axial current matrix element, interpreted as
the portion of the total spin carried by the quarks $\Delta \Sigma _{{\rm {p}%
}}=a_{0}$, is much smaller than $1$.

Polarized structure functions have been thoroughly investigated within the
perturbative QCD (see for review \cite{AEL} and recent papers \cite
{Cheng,Goto}). In the present note, following Ref.~\cite{KarLip} we shall
use $I_{{\rm {p}}}=0.124\pm 0.011$ which can be translated into:
\begin{equation}
\Gamma _{{\rm {p}}}\equiv 4\Delta u_{{\rm {p}}}+\Delta d_{{\rm {p}}}+\Delta
s_{{\rm {p}}}=2.56\pm 0.23\,.  \label{Gampval}
\end{equation}
if $\alpha _{{\rm s}}(Q^{2}=3~({\rm GeV}/c)^{2})=0.4$. Then, assuming the
Bjorken sum rule, one gets for the neutron
\begin{equation}
\Gamma _{\rm n}\equiv 4\Delta d_{{\rm {p}}}+\Delta u_{{\rm {p}}}+\Delta s_{{\rm {%
p}}}=-0.928\pm 0.186\,.  \label{Gamnval}
\end{equation}

It is important to realize that $\Delta \Sigma _{{\rm {p}}}$ is {\em not
directly measured}; it is extracted from the data through some theoretical
model. If the SU(3) symmetry is assumed then all hyperon semileptonic decays
can be expressed in terms of $2$ constants $F$ and $D$ or alternatively in
terms of two axial charges
\begin{eqnarray}
g_{A} &=&\Delta u_{{\rm {p}}}-\Delta d_{{\rm {p}}}=F+D,  \nonumber \\
a_{8} &=&\Delta u_{{\rm {p}}}+\Delta d_{{\rm {p}}}-2\Delta s_{{\rm {p}}%
}=3F-D.  \label{8currents}
\end{eqnarray}
The SU(3) symmetry alone does not relate the singlet current to the octet
ones, therefore in order to extract $a_{0}$ an extra input is needed, for
example one can use $\Gamma _{{\rm {p}}}$ of Eq.(\ref{Gampval}):
\begin{equation}
a_{0}=\Delta \Sigma =\frac{1}{2}\left( \Gamma _{{\rm p}}-3F-D\right) .
\label{0current}
\end{equation}
In order to extract constants $F$ and $D$ one usually uses the neutron beta
decay and $\Sigma^{-}\rightarrow {\rm n}$ decay which gives the so called 
{\em typical }SU(3) values: $F=0.46$, $D=0.8$ and $a_{8}=0.58$.
Alternatively one can perform the least $\chi ^{2}$ fit to {\em
all} hyperon decays. Such a fit was performed for example in
Ref.\cite{Man}, with the result $F=0.47\pm 0.01$, $D=0.79\pm 0.01$
and with relatively large $\chi ^{2}=13.5$ for $4$ degrees of
freedom which is an indication of the SU(3) breaking. The
resulting $a_{8}=0.62\pm 0.03$. It has been noticed in \cite{Man}
that nominal errors obtained from the fit are underestimates of
the true ones.

\begin{wrapfigure}{l}{8cm}
\epsfig{figure=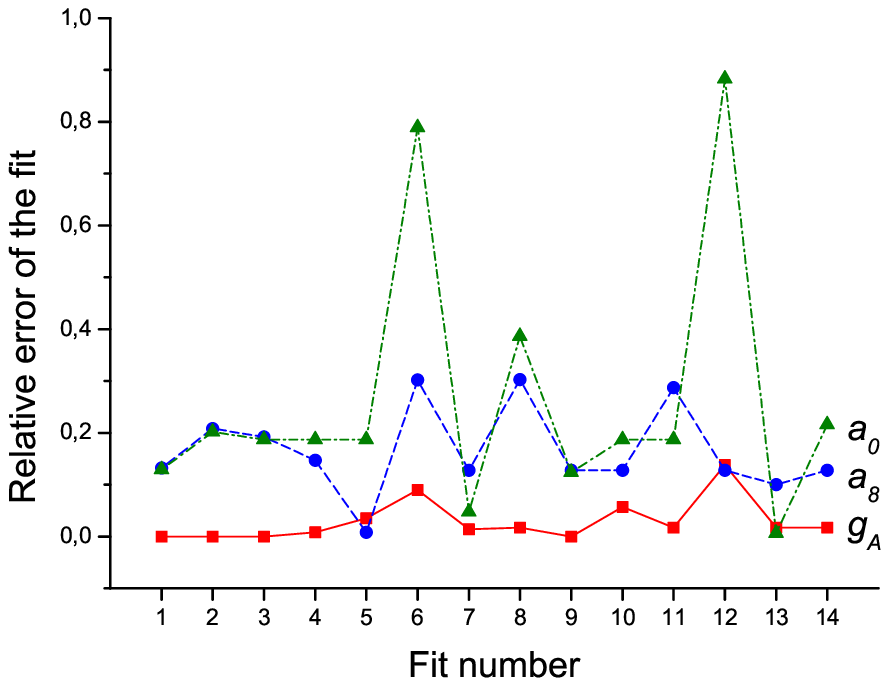,width=8cm}\\
{\small Figure 1: Relative errors of axial charges for 14
different fits described in the text (see also Fig.2). }
\medskip
\end{wrapfigure}

It is instructive to perform an oversimplified but illustrative analysis of
the data which leads to the same conclusion. If the SU(3) symmetry breaking
was not important, any pair out of six known semileptonic decays should give
roughly the same number for $g_{A}$, $a_{8}$ and $a_{0}$. This is, however,
{\em not} the case. There are $14$ pairs which one can form out of 6 known
hyperon decays which give linearly independent equations for $g_{A}$, $a_{8}$
and $a_{0}$ (with $\Gamma _{{\rm {p}}}$ of Eq.(\ref{Gampval}) as an extra
input for the latter). Averaging over these $14$ possibilities (disregarding
the experimental errors) one gets
\begin{equation}
\bar{g}_{A}=1.27,\quad \bar{a}_{8}=0.67,\quad \bar{a}_{0}=0.17.
\end{equation}
We see that $\bar{g}_{A}$ is almost identical to the $g_{A}$ measured in the
$\beta $ decay. In Fig.1 we plot the relative error for each fit for $g_{A}$%
, $a_{8}$ and $a_{0}$. The errors of $g_{A}$ are by far the smallest and the
largest variations are observed for $a_{0}$. This can be further
demonstrated by averaging the relative error over the $14$ fits. Then
\begin{equation}
\overline{\Delta g_{A}}=0.03<\overline{\Delta a_{8}}=0.17<\overline{\Delta
a_{0}}=0.27.  \label{hier}
\end{equation}
If a similar analysis is performed for the baryon masses in the
SU(3) octet and decuplet then one gets that the average relative
mass error is of the order of 10\% which may be interpreted as a
{\em typical} accuracy of the SU(3) symmetry for baryons.
Therefore an interesting pattern of the SU(3) symmetry breaking
for the hyperon decays can be seen in the data: $g_{A}$ remains
almost unaffected (which may be interpreted as the sign of the
almost exact isospin symmetry) whereas for $a_{8}$ the SU(3)
symmetry breaking is {\em typical} (17\% error). For $a_{0}$ the
average error is, however, larger, which is the signature of a
potentially large symmetry breaking in this channel. This pattern
can be in principle understood theoretically since in the symmetry
limit $g_{A}$ is given as a sum of two positive constants
(\ref{8currents}) $F$ and $D$, and therefore the theoretical error
for $g_{A}$ is of the order of the typical error of $F$ and $D$
alone, whereas $a_{8}$ is given as a {\em difference}
(\ref{8currents}) and this of
course increases the error\footnote{%
One has to keep in mind, however, that this is just a rough
argument since the errors of $F$ and $D$ are correlated.}. The
same argument applies in principle also to $a_{0}$, however here
the situation is less clear, since
we have to deal with 3 quantities (\ref{0current}) $F$, $D$ and $\Gamma _{%
\text{p}}$ not just $2$.

\begin{wrapfigure}{l}{8cm}
\epsfig{figure=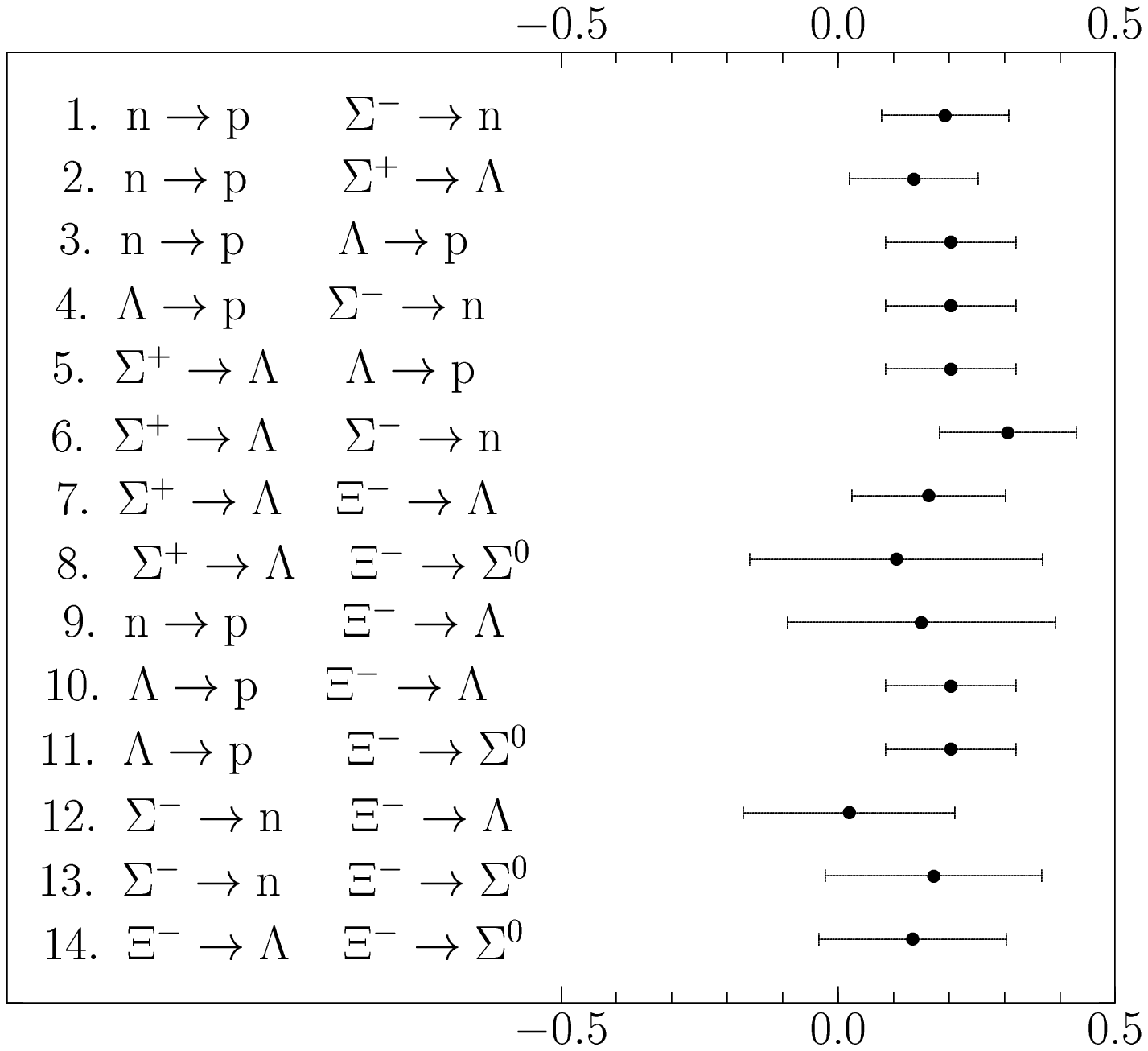,width=7.5cm}\\
{\small Figure 2: Values of $a_0$ for 14 different inputs with experimental errors.}
\medskip
\end{wrapfigure}

In Fig.2 we plot 14 values of $a_{0}=\Delta \Sigma _{{\rm {p}}}$ together
with the experimental error bars. Typically the large error bars are due to
the large error of $\Gamma _{{\rm {p}}}$ and of the semileptonic $\Xi ^{-}$
decays.

The above analysis suggests that the good measure of the SU(3)
symmetry breaking is the value of $a_{8}$. One may ask if the
polarized DIS data can differentiate between different values of
$a_{8}$? This question was addressed in Refs.\cite{LSS}, where the
constrained fit to all existing DIS data
has been performed for 3 different values of $a_{8}$: $0.40,$ $0.58$ ({\em %
typical} SU(3) value) and $0.86$. Unfortunately the $\chi ^{2}$ value for
all 3 fits is practically the same indicating that the present DIS data are
not able to decide in which way the SU(3) symmetry is broken. In Fig.3 we
plot the values of $\Delta q$, $\Delta G$ and $\Delta \Sigma $ from Ref.\cite{LSS}
(in JET renormalization scheme)
as functions of $a_{8}$ together with the extrapolation to even smaller
values of $a_{8}$, not considered in \cite{LSS}
which were, however, suggested in Ref. \cite{Man}.

\begin{wrapfigure}{l}{8cm}
\epsfig{figure=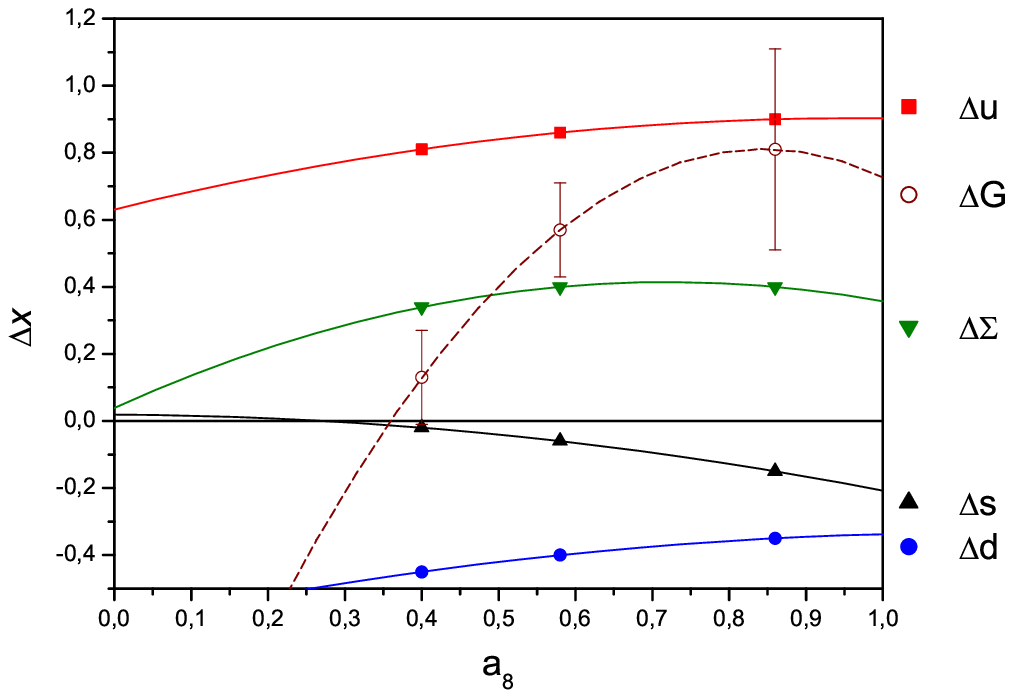,width=8cm}\\
{\small Figure 3: $\Delta q$, $\Delta G$ and $\Delta\Sigma$
for the proton  from
Ref.\cite{LSS} as functions of $a_8$ together with
the extrapolation.}
\medskip
\end{wrapfigure}

In the situation where the DIS data are not able to provide any hint how the
SU(3) symmetry is broken, one has to resort to some specific models~\cite
{KarLip}. In this paper, following Refs.\cite{KimPraGo2}, we will use the
Chiral Quark-Soliton Model ($\chi $QSM for short)~\cite{DPP,WakamatsuYoshiki}
(see Ref.\cite{review} for review) to implement {\em perturbatively} the
symmetry breaking due to the non-zero strange quark mass. This model
satisfactorily describes the axial-vector properties of the hyperons \cite
{BloPraGo}\nocite{BPG,Wakaspin}--\cite{KimPoPraGo}. Since the symmetry
breaking pattern of the $\chi $QSM is identical to the one derived in large $%
N_{{\rm c}}$ QCD \cite{Man}, our analysis is in fact much more general than
the model itself.

The $\chi $QSM (as most of the hedgehog models~\cite{SchechterWeigel}) has a
remarkable virtue of connecting the singlet axial-vector constant $a_{0}$
with $g_{{\rm A}}$ and $a_{8}$ in a direct manner. This connection
introduces a model dependence into our analysis. However, as we discussed in
\cite{KimPraGo2}, there is no significant numerical difference between the
results obtained with and without this model dependent ingredient. Our
analysis is based on a ''{\em model-independent}'' method where the
dynamical quantities, which are in principle calculable within the model
\cite{BloPraGo}, are treated as free parameters. By adjusting them to the
experimentally known semileptonic decays we allow not only for maximal
phenomenological input but also for minimal model dependence. In Refs.\cite
{KimPraGo}\nocite{strange,Min}-- \cite{HongPark} magnetic moments of the
octet and decuplet have been studied in this way. Model calculations for the
vector-axial properties of baryons have been presented in Ref.\cite
{KimPoPraGo}. There exist also direct model calculations of the spin
polarization function itself~\cite{WakaWata,Goekespin,PolDu}.

The transition matrix elements of the hadronic axial-vector current $\langle
B_{2}|A_{\mu }^{X}|B_{1}\rangle $ can be expressed in terms of three
independent form factors:
\begin{equation}
\langle B_{2}|A_{\mu }^{X}|B_{1}\rangle \;=\;\bar{u}_{B_{2}}(p_{2})\left[
\left\{ g_{1}(q^{2})\gamma _{\mu }-\frac{ig_{2}(q^{2})}{M_{1}}\sigma _{\mu
\nu }q^{\nu }+\frac{g_{3}(q^{2})}{M_{1}}q_{\mu }\right\} \gamma _{5}\right]
u_{B_{1}}(p_{1}),
\end{equation}
where the axial-vector current is defined as
\begin{equation}
A_{\mu }^{X}\;=\;\bar{\psi}(x)\gamma _{\mu }\gamma _{5}\lambda _{X}\psi (x)
\label{Eq:current}
\end{equation}
with $X=\frac{1}{2}(1\pm i2)$ for strangeness conserving $\Delta S=0$
currents and $X=\frac{1}{2}(4\pm i5)$ for $|\Delta S|=1$.

The $q^{2}=-Q^{2}$ stands for the square of the momentum transfer $%
q=p_{2}-p_{1}$. The form factors $g_{i}$ are real and depend only on $Q^{2}$
in the case of the $CP$-invariant processes. We will neglect here both $g_{3}$
and $g_{2}$. In principle the latter form factor is proportional to $m_{{\rm %
s}}$ and therefore should be included in the consistent analysis
of the weak decays data. Unfortunately, such an analysis is still
missing and all experimental results for $g_{1}$ assume
$g_{2}\equiv 0$ (with an exception of $\Sigma ^{-}$ decay
\cite{Hsueh}).

Other possible small $m_{{\rm s}}$ corrections come from the evolution of $%
g_{1}$ with $Q^{2}$, due to the non-conservation of the axial-vector
currents caused by the SU(3) symmetry breaking. These corrections are also
neglected in our approach.

The $\chi $QSM allows to express all semileptonic hyperon decays in terms of
6 parameters. These include both symmetry breaking in the baryon wave
functions and in the weak currents as well. Unfortunately this is precisely
the number of the known amplitudes $A_{i}$ of the semileptonic weak decays:
\begin{equation}
\begin{array}{ccc}
A_{1}=(g_{1}/f_{1})^{({\rm n}\rightarrow {\rm p})}, & A_{3}=(g_{1}/f_{1})^{(%
\Lambda \rightarrow {\rm p})}, & A_{5}=(g_{1}/f_{1})^{(\Xi ^{-}\rightarrow
\Lambda )}, \\
A_{2}=(g_{1}/f_{1})^{(\Sigma ^{\pm }\rightarrow \Lambda )}, &
A_{4}=(g_{1}/f_{1})^{(\Sigma ^{-}\rightarrow {\rm n})}, &
A_{6}=(g_{1}/f_{1})^{(\Xi ^{-}\rightarrow \Sigma ^{0})}.
\end{array}
\label{Ai}
\end{equation}
The U(3) axial-vector constants $g_{A}^{(0,3,8)}$ can be also expressed in
terms of the same set of parameters.

It is important to notice that there exist two linear combinations of $A_{i}$%
's which are free of the linear $m_{{\rm {s}}}$ corrections in the $\chi $%
QSM (and large $N_{{\rm c}}$ QCD \cite{Man}), namely:
\begin{eqnarray}
F &=&\frac{1}{12}(4A_{1}-4A_{2}-3A_{3}+3A_{4}+3A_{5}+5A_{6}),  \nonumber \\
D &=&\frac{1}{12}(4A_{2}+3A_{3}-3A_{4}-3A_{5}+3A_{6})  \label{DFexp}
\end{eqnarray}
which give numerically
\begin{equation}
F=0.50\pm 0.07\;~~{\rm and}\;~~D=0.77\pm 0.04.  \label{FDmsfree}
\end{equation}
We shall employ these values of $F$ and $D$ for the fits in the symmetry
limit. Note that in this case $a_{8}=0.73$ which is higher than the {\em %
typical} symmetry value $0.58$ and the least $\chi ^{2}$ value
$0.62$. This shows that the value of $a_{8}$ depends on the way of
analyzing the data even in the symmetry limit.

In the following we shall present two sets of results: 1) the SU(3)
symmetric ones with the input values of $F$ and $D$ given by Eq.(\ref
{FDmsfree}) and 2) the ones with the symmetry breaking included where all 6
free parameters are fitted to the 6 hyperon decays of Eqs.(\ref{Ai}).
Interestingly, the model predicts $(g_{1}/f_{1})^{(\Xi ^{-}\rightarrow
\Sigma ^{0})}=(g_{1}/f_{1})^{(\Xi ^{0}\rightarrow \Sigma ^{+})}=1.278\pm
0.158$ in very good agreement with the recent result for the latter decay
from the KTeV collaboration \cite{KTeV} $(g_{1}/f_{1})^{(\Xi ^{0}\rightarrow
\Sigma ^{+})}=1.32_{-0.17}^{+0.21}\pm 0.05$ (first error is statistical,
second systematic). Unfortunately this particular decay is not a good
measure of the SU(3) breaking since the equality of the two decays holds
both in the symmetry limit and with the breaking included.

In the SU(3) symmetry limit the model predicts that
\begin{equation}
\Delta \Sigma =9F-5D  \label{9F5D}
\end{equation}
for all octet baryons. This formula has a remarkable feature: it
interpolates between the naive quark model and the Skyrme model. Indeed, in
the case of the naive quark model where $F=2/3$ and $D=1$ one obtains $%
\Delta \Sigma =1$, whereas in the case of the simplest Skyrme
model for which $F/D=5/9$, $\Delta \Sigma =0$, as observed for the
first time in Ref. \cite{BroEllKar}. Notice that $\Delta \Sigma $
is very sensitive to small variations of $F$ and $D$, since it is
a difference of the two, with relatively large coefficients.
Indeed, for the 14 fits of Figs.1 and 2 the central value for
$\Delta \Sigma $ varies between $-0.25$ to approximately 1. This
might be the likely explanation of the hierarchy discussed in
connection with Eq.(\ref{hier}).

\centerline{\epsfysize=2.3in\epsffile{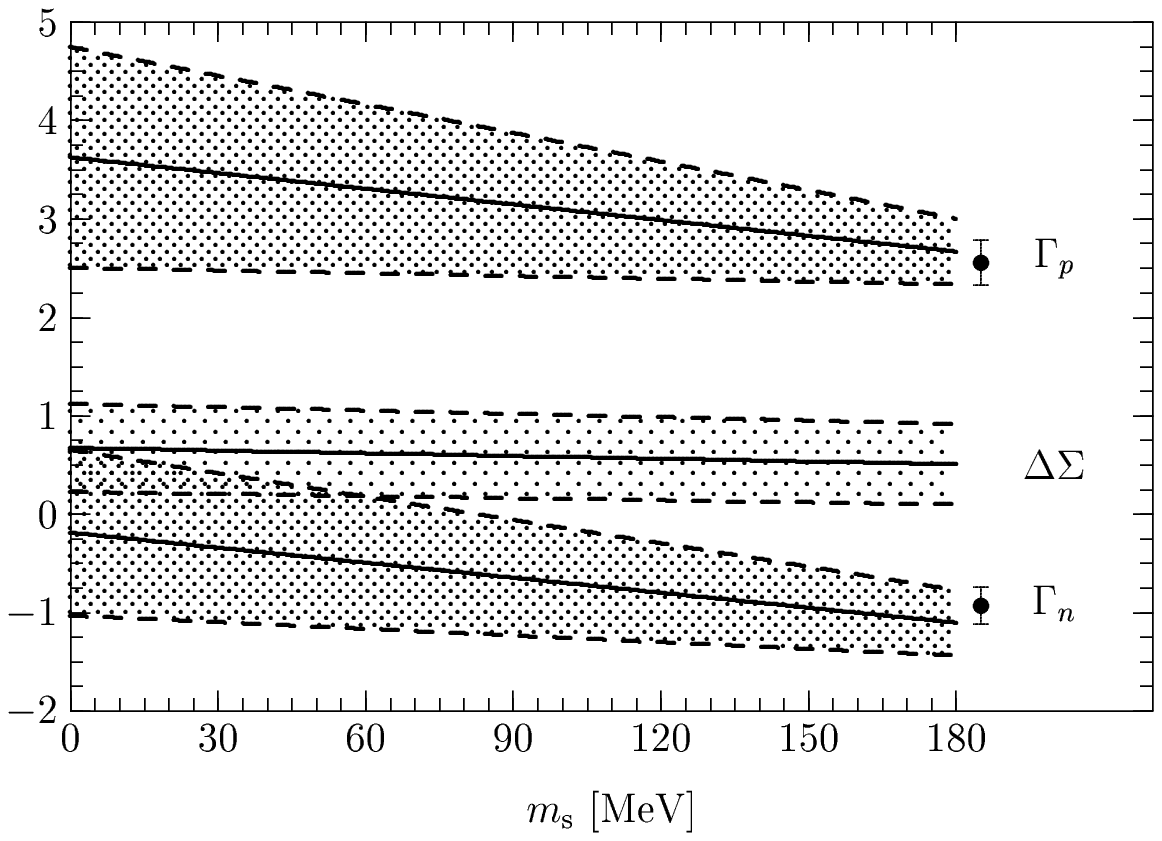}
\epsfysize=2.3in\epsffile{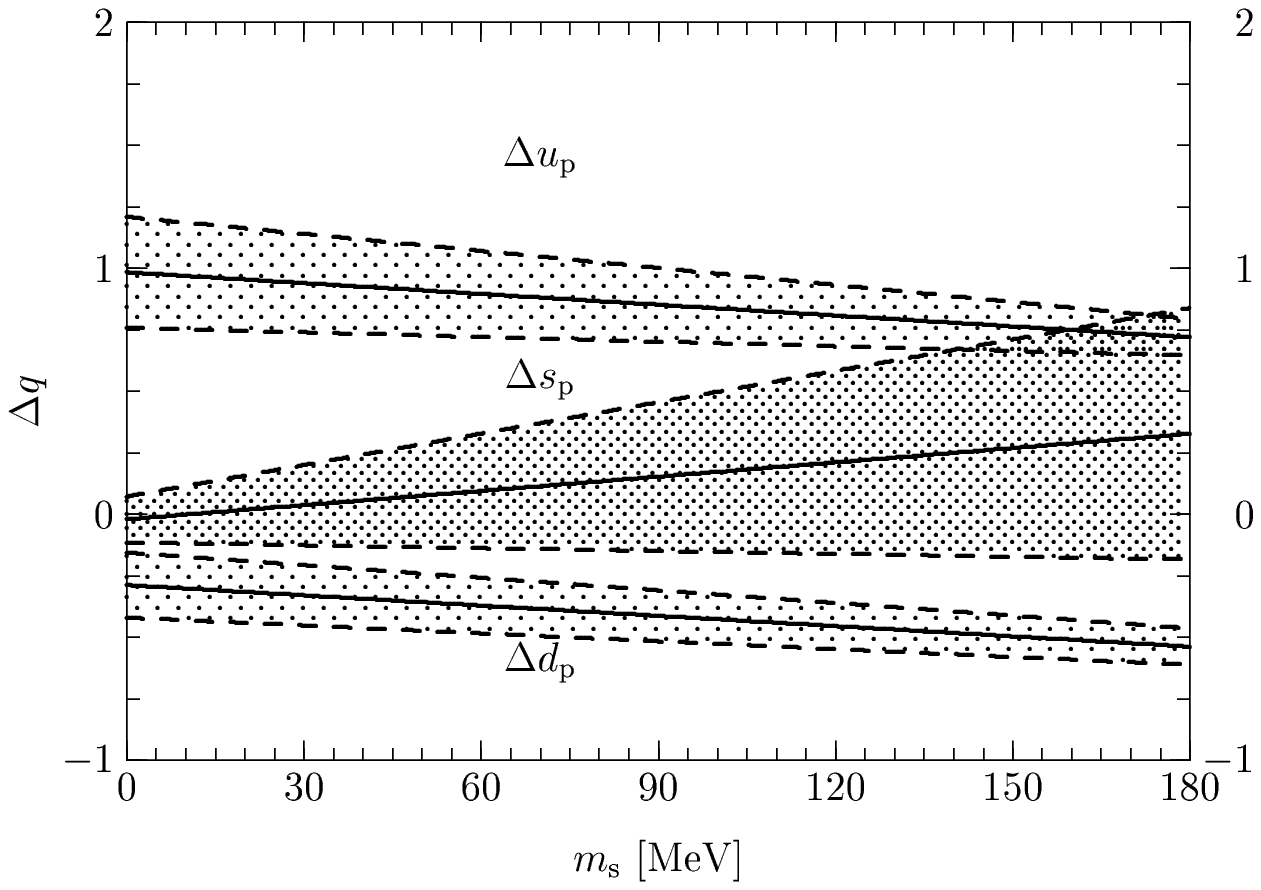}} 
{\small Figure 4: Figure $\Gamma _{{\rm p}}$, $\Gamma _{{\rm n}}$, $\Delta
\Sigma $ and $\Delta q$ for the proton in the SU(3) limit and with SU(3)
symmetry breaking, interpolated according to Eq.(13), together with the
error bands.} \medskip 

In the plots of $\Delta q$'s we have restored the linear $m_{{\rm {s}}}$
dependence in the following way:
\begin{equation}
\Delta q=\Delta q^{(0)}+\frac{m_{{\rm {s}}}}{180\,{\rm {MeV}}}\left( \Delta
q-\Delta q^{(0)}\right) ,  \label{msdep}
\end{equation}
assuming for definiteness that $m_{{\rm {s}}}=180$ MeV. This is
possible because our chiral parameters $F$ and $D$ need not to
be refitted as the symmetry breaking corrections are included.

In Fig.4 we plot $\Gamma _{\text{p,n}}$ and $\Delta \Sigma _{\text{p}}$ both
for the chiral symmetry fit and for the full fit together with experimental
data for the proton and neutron. We see that in the chiral limit defined by (%
\ref{FDmsfree}) $\Gamma _{\text{p}}$ and $\Gamma _{\text{n}}$ are not
reproduced. Only if the symmetry breaking terms are included, the model
predictions hit the experimental values. Unfortunately our predictions have
large error due to the experimental error of $\Xi ^{-}$ decays. Somewhat
unexpectedly we see that $\Delta \Sigma _{\text{p}}$ is almost independent
of the chiral symmetry breaking\footnote{%
Similar behavior has been observed in Ref.\cite{LichtLip}.} and stays within
the range $0.1\rightarrow 1.1$, if the errors of the hyperon decays are
taken into account. Similarly to $\Gamma _{\text{n,p}}$, $75\%$ of the
experimental error of $\Delta \Sigma _{\text{p}}$ comes from the two least
known hyperon decays $\Xi ^{-}\rightarrow \Lambda ,\,\Sigma ^{0}$
(corresponding to $A_{5}$ and $A_{6}$).

\begin{wrapfigure}{l}{8cm}
\epsfig{figure=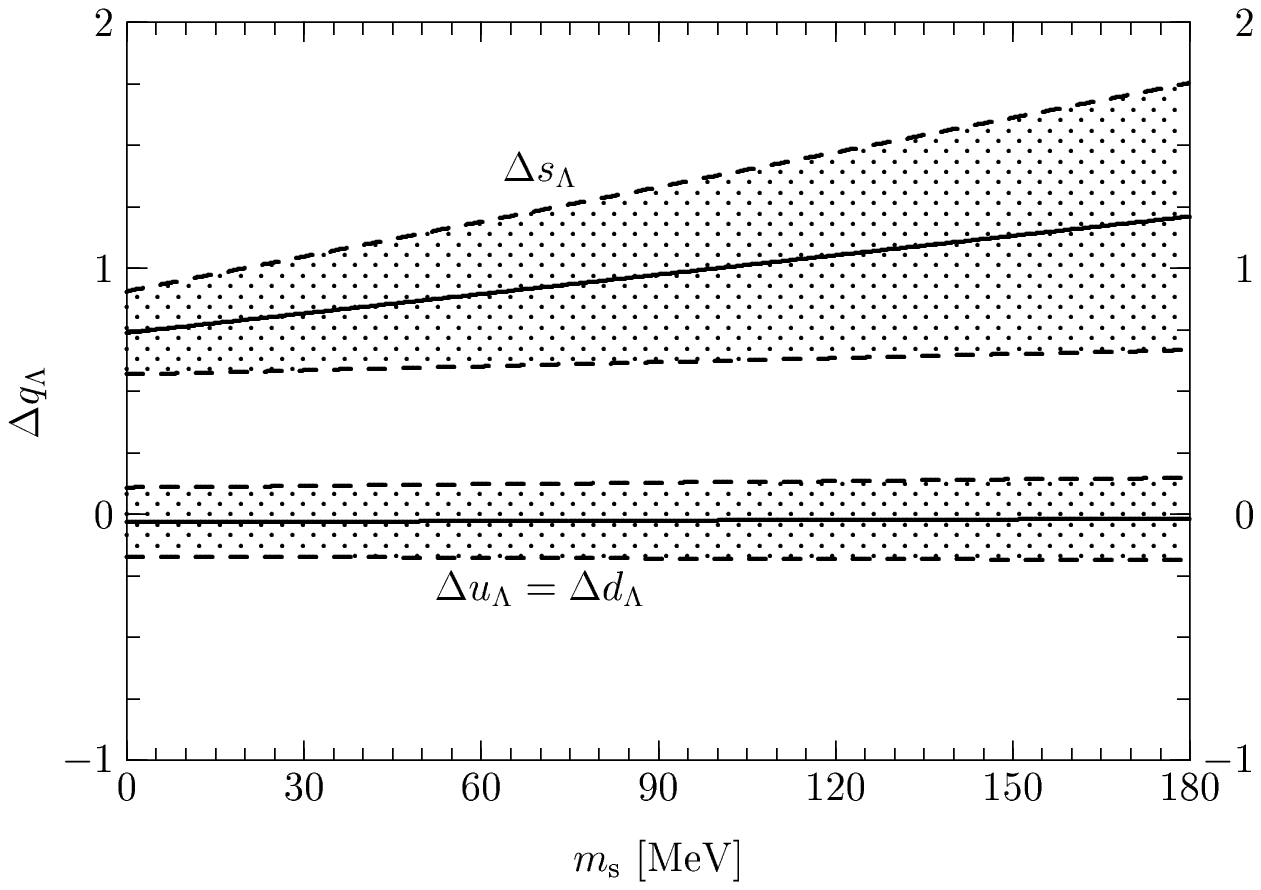,width=8cm}\\
{\small Figure 5: Same as Fig.4 for $\Lambda$ hyperon. }
\bigskip\bigskip

\epsfig{figure=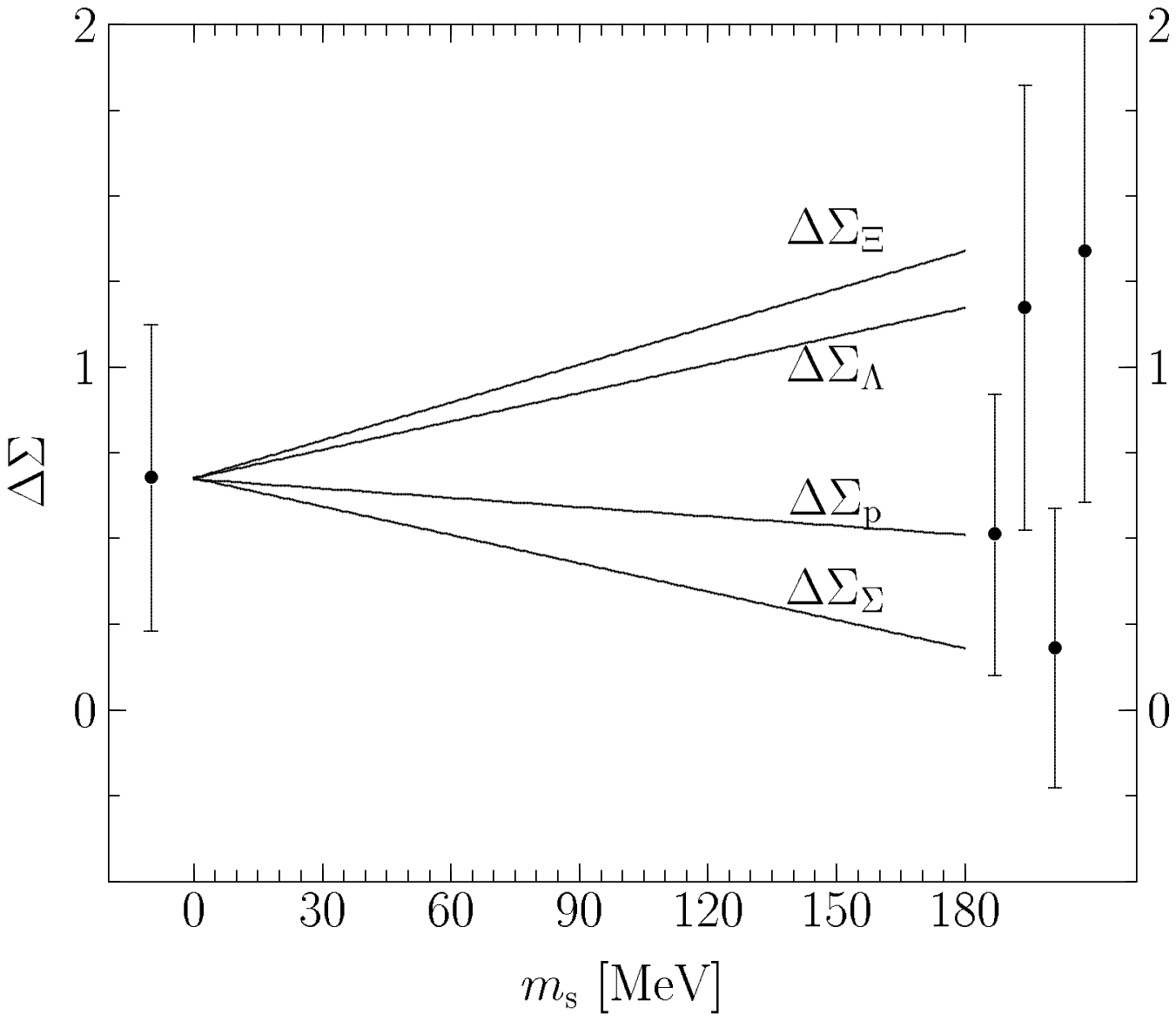,width=7.5cm,height=5.5cm}\\
{\small Figure 6: $\Delta\Sigma$ for octet baryons without and
with symmetry breaking together with the errors (described in the
text). }
\medskip
\end{wrapfigure}

The advantage of the present approach is that it can be easily extended to
all hyperons as it was done in Refs.\cite{KimPraGo2}. Unfortunately there
is very little
hope that $\Delta q$'s in hyperons can be measured. Nevertheless, one can
make use of an approximate equality of the quark distributions and the
fragmentation functions \cite{GriLip} (at least for large $x$) which are
accessible experimentally. Here of special interest is the $\Lambda $
hyperon for which there exist experimental data.
In Fig.5. we plot $\Delta q_{\Lambda }$
both in the SU(3) symmetry limit and with the symmetry breaking.
Interestingly $\Delta u_{\Lambda }=\Delta d_{\Lambda }$ is compatible with
zero in both limits, whereas $\Delta s_{\Lambda }$ increases with $m_{{\rm {s%
}}}$ (with large errors). This increase is due to the fact that the singlet
axial-vector currents $\Delta \Sigma $'s split when the symmetry breaking is
switched on. This splitting is depicted in Fig.6. We see that $\Delta \Sigma
_{{\rm {p}}}$ shows the weakest $m_{{\rm {s}}}$ dependence, whereas $\Delta
\Sigma _{\Lambda }$ and $\Delta \Sigma _{\Xi }$ depend quite strongly on $m_{%
{\rm {s}}}$. Large error bars for these quantities are due almost entirely
to the large errors of $\Xi^{-} $ decays $A_{5}$ and $A_{6}$. This feature that $%
\Delta \Sigma $'s are different for different baryons is characteristic for
the models where the symmetry breaking in the currents is included. If the
symmetry breaking is taken into account only in the wave functions then $%
\Delta \Sigma $'s remain identical for all baryons \cite{WeiSig}.

As it was suggested in the introduction a good measure of the SU(3) symmetry
breaking is the value of $a_{8}$. In our model $a_{8}$ is basically
undetermined since $a_{8}=-0.47\pm 1.14$. Here $88\%$ of the error comes
from the decay $\Xi ^{-}\rightarrow \Sigma ^{0}$. One may, however, conclude
that $a_{8}$ is smaller than our symmetry limit value $a_{8}=0.73$ and
perhaps smaller than the {\em typical} SU(3) value $0.58$. In the similar
model \cite{WeiSig}, where, however, the symmetry breaking is included
only in
the wave functions and the exact diagonalization procedure is applied, $%
a_{8}=0.3-0.4$ depending on the fit, $\Delta \Sigma _{\text{p}%
}=0.27-0.29$ and $\Delta s_{\text{p}}=-0.02$ to $-0.034$, a fairly small
number. This shows that the Chiral Quark Soliton Models, independently of
the details, agree well with large $N_{c}$ QCD fit (with the SU(3) breaking
included) \cite{Man} to the existing data (including decuplet decays), which
gives $a_{8}=0.27\pm 0.09$. A simple parametrization of the SU(3) breaking
proposed in Ref.\cite{EhrSch} assumed that the {\em effective} ratio $F/D$ for
given decay $i\rightarrow f$ was proportional to the dimensionless parameter
$\delta =\left[ \left( m_{i}+m_{f}\right) -\left( m_{p}+m_{n}\right) \right]
/\left[ \left( m_{i}+m_{f}\right) +\left( m_{p}+m_{n}\right) \right] $.
Then, by an appropriate interpolation, for the neutron $\beta $ decay one
gets $F/D=0.49\pm 0.08$, less than the {\em typical} SU(3) value. This
translates in turn into $a_{8}=0.4$. Similar value for $a_{8}$ was obtained
in Ref.\cite{LichtLip} where a simple model of strange s\={s}-pairs suppression,
parametrized by a single parameter $\varepsilon $, both in neutron and $%
\Sigma ^{-}$, was considered. The resulting value of $\Delta \Sigma _{\text{p%
}}=0.28-0.31$ was almost independent of $\varepsilon $, whereas $\Delta s_{%
\text{p}}$ was quite sensitive, {\em increasing} with $\varepsilon $ from $%
-0.10$ at $\varepsilon =0$ to $-0.04$ at $\varepsilon =2$. In this
respect the model of Ref.\cite{LichtLip} is in qualitative
agreement with our present study. Subsequently $a_{8}$ calculated
from Eq.(11) of Ref.\cite{LichtLip}
{\em %
decreases} from the symmetry value $a_{8}=0.575$ to $a_{8}=0.42$ for $%
\varepsilon =2$. Interestingly, for this range of {\em small}
$a_{8}$ the
global analysis of Ref.\cite{LSS} indicates, as seen from Fig.3, that $\Delta s_{%
\text{p}}$ is compatible with zero and $\Delta G$ is also very small.

There are, however, other models which predict $a_{8}$ {\em larger} than the
{\em typical} symmetry value. Chiral perturbation theory with kaon loops
\cite{SavWal} gives $a_{8}=0.8$. Here both $\Delta s_{\text{p}}$ and $%
\Delta \Sigma _{\text{p}}$ are sensitive to the strength of the
symmetry breaking, while $\Delta u_{\text{p}}$ and $\Delta
d_{\text{p}}$ are not. The successful phenomenological model of
S.B.Gerasimov \cite{SBG} presented at this conference predicts in
general $a_{8}>0.8$. An even larger value for $a_{8}$ was obtained
in Ref.\cite{LinSnell} in a model where quark spin polarization is
mass dependent.

To complete the model spectrum let us mention the recoil model of
Ref.\cite{Rat} where the ratio $F/D$ (hence $a_{8}$) remains not affected
by the symmetry breaking. However, Ref.\cite{Rat} predicted $%
(g_{1}/f_{1})^{(\Xi ^{0}\rightarrow \Sigma ^{+})}=1.17$ to $1.14$ 
$\pm 0.03$
which is significantly lower than the recentKTeV result \cite{KTeV}. 
Similar value for $%
a_{8}$ was obtained in Ref.\cite{SKM} where the most general form
of a baryon matrix element of an axial-vector current has been
considered. In this respect the model of Ref.\cite{SKM} is in
principle parallel to our approach; also the number of free
parameters is $6$ ($4$ of them proportional to $m_{s}$) as in our
case. However, the authors of Ref.\cite{SKM} did not consider two
measured decays $\Sigma ^{+}\rightarrow \Lambda $ and $\Xi
^{-}\rightarrow \Sigma ^{0}$. For that reason they tried to reduce
the parameter space and performed 4 fits with only one $m_{s}$
dependent parameter being different from 0. This is perhaps the
reason why their conclusions are different from ours.

Let us briefly summarize our findings. We have performed the ''{\em %
model-independent}'' analysis of the hyperon semileptonic decays based on
the algebraic structure of the Chiral Quark Soliton Model. There are two
model ingredients which are of importance. The first one is the model
formula for the octet axial-vector currents which have been derived in the
linear order in $m_{{\rm s}}$ and $1/N_{{\rm c}}$. Our formulae here have
the same algebraical structure as in the large $N_{{\rm c}}$ QCD \cite{Man},
and therefore they are more general than the model itself. Secondly, in
contrast to pure QCD, the model provides a link between the octet
axial-vector currents and the singlet axial-vector current. This connection
is a truly model-dependent ingredient, however, we have given the arguments
in favor of Eq.(\ref{9F5D}) based on the fact that apart from the general
success of the $\chi $QSM in reproducing the form factors and parton
distributions, in the limit of the small soliton it properly reduces to the
Nonrelativistic Quark Model prediction, and in the limit of the large
soliton it reproduces the Skyrme Model prediction for $\Delta \Sigma $.
Similarly, in Ref.\cite{WeiSig} the argument has been given that Eq.(\ref
{9F5D}) naturally emerges in the limit of the large $m_{{\rm s}}$, where the
SU(3) flavor symmetry reduces to the SU(2) one. In Refs.\cite{KimPraGo2} 
we have
shown that if we abandon the model formula for $a_{0}$ and use instead $%
\Gamma _{{\rm {p}}}$ as an additional input, then the resulting $a_{0}$ is
numerically almost identical to the one obtained within the model. This
provides a further support for the model formula for $a_{0}$.

Numerically our results suffer from large errors due to the
experimental errors of the $\Xi ^{-}$ decays. It is therefore of
utmost importance to measure these two decays (and also $\Xi
^{0}\rightarrow \Sigma ^{+}$) with the precision comparable to the
other four decays. One should bare in mind that this is one of a
few cases, where the low energy data have an important impact on
our understanding of the high energy scattering. Given the
theoretical implications of these experiments as far as the role
of the axial anomaly and the gluon polarization is concerned
\cite{AEL,Cheng,Goto}, one should make it clear how important the
new measurements of the $\Xi ^{-,0}$ decays would be.

Within the experimental uncertainty we conclude that in our approach $a_{8}$
is smaller than the {\em typical} SU(3) value, and $\Delta s_{\text{p}}$ in
the proton is not incompatible with $0$. Another important feature is that $%
\Delta \Sigma $'s for different hyperons split when the symmetry breaking is
included. One should, however, keep in mind that the theoretical situation
concerning the size and direction of the SU(3) breaking is not clear.

This work was supported by the Polish KBN Grant PB~2~P03B~019~17 
and Bogolyubov-Infeld Program of JINR-Poland collaboration (M.P.),
the Korea Science \& Engineering Foundation
grant  No. R01-2001-00014 (H.-Ch.K.) and the BMBF, DFG, and COSY--Project.
M.P. thanks the organizers of the Spin 2001 Workshop, especially
A.V. Efremov, form warm hospitality
and B.S. Gerasimov, A.V. Sidorov, D.B. Stamenov and O.V. Teraev for
discussion and remarks.


\end{document}